\documentclass[letterpaper,american,aps, amsfonts, amssymb, amsmath, reprint, showkeys, nofootinbib, superscriptaddress,  prb, citeautoscript, longbibliography]{revtex4-1}
\usepackage{lmodern}
\usepackage[T1]{fontenc}
\usepackage[latin9]{inputenc}
\usepackage{color}
\usepackage{amsmath}
\usepackage{amssymb}
\usepackage{graphicx}
\usepackage[dvipsnames]{xcolor}

\makeatletter

\newcommand{\noun}[1]{\textsc{#1}}
\makeatother

\usepackage{babel}
\begin{document}
\title{Identifying Majorana vortex modes via non-local transport}
\author{Bj\"orn Sbierski}
\affiliation{Department of Physics, University of California, Berkeley, California
94720, USA}
\affiliation{Department of Physics and Arnold Sommerfeld Center for Theoretical
Physics, Ludwig-Maximilians-Universit\"at M\"unchen, Theresienstr.~37,
D-80333 M\"unchen, Germany }
\author{Max Geier}
\affiliation{Center for Quantum Devices, Niels Bohr Institute, University of Copenhagen,
DK-2100 Copenhagen, Denmark}
\author{An-Ping Li}
\affiliation{Center for Nanophase Materials Sciences, Oak Ridge National Laboratory, Oak Ridge, TN 37831, USA}
\author{Matthew Brahlek}
\affiliation{Materials Sciences and Technology Division, Oak Ridge National Laboratory, Oak Ridge, TN 37831, USA}
\author{Robert G. Moore}
\affiliation{Materials Sciences and Technology Division, Oak Ridge National Laboratory, Oak Ridge, TN 37831, USA}
\author{Joel E. Moore}
\affiliation{Department of Physics, University of California, Berkeley, California
94720, USA}
\affiliation{Materials Sciences Division, Lawrence Berkeley National Laboratory,
Berkeley, California 94720, USA}
\date{\today}
\begin{abstract}
The combination of two-dimensional Dirac surface states with s-wave superconductivity is expected to generate localized topological Majorana zero modes in vortex cores. Putative experimental signatures of these modes have been reported for heterostructures of proximitized topological insulators, iron-based superconductors
or certain transition metal dichalcogenides. Despite these efforts, the Majorana nature of the observed excitation is still under debate. We propose to identify the presence of Majorana vortex
modes using a non-local transport measurement protocol originally employed for one-dimensional settings. In the case of an isolated subgap state, the protocol provides a spatial map of the ratio of local charge- and probability-density which offers a clear distinction between Majorana and ordinary fermionic modes. We show that these distinctive features survive in the experimentally relevant case of hybridizing vortex core modes.
\end{abstract}
\maketitle

\section{Introduction\label{sec:Introduction}}

In condensed matter physics, Majorana zero energy modes are highly sought
after subgap states localized in topological superconductors and certain fractional quantum Hall states\citep{Alicea2012,Sau2021}.
Whereas initial efforts were mainly directed towards one-dimensional
systems based on semiconductor quantum wires in proximity with conventional
superconductors (``Majorana''-wire)\citep{Mourik2012}, recent progress
in this direction has been slowed by the ambiguity related to the
interpretation of transport measurements \citep{Zhang2021a} local
to the ends of the wire. As a consequence, it has been proposed that
non-local transport setups can give a much cleaner picture of
the nature of subgap states\citep{Rosdahl2018,Danon2020,PanThreeTerminal2021,Pikulin2021}
with a small number of recent experiments already available \citep{Menard2020,Puglia2020}.

Candidate systems for Majorana zero modes are not limited
to one spatial dimension. In a classic paper\citep{Fu2008}, Fu and Kane proposed
to realize Majorana zero modes in the center of a vortex in the superconducting
order parameter assuming the latter pairs a single-species of two-dimensional Dirac quasiparticles. The resulting zero-energy excitations are also known as Majorana vortex
modes (MVM). The initial proposal was framed in the context of topological insulator surface states proximitized to a superconducting layer, which was subsequently realized in experiment\cite{Xu2015b,Sun2016}. However the unambiguous identification of MVM in the experimentally observed local density of states (LDOS) is complicated by the fact that the putative MVM at $E_0=0$ is by far not the only subgap state localized at the vortex position. In addition, theory predicts a whole ladder of finite-energy
Caroli-de Gennes-Matricon (CdGM) states, with
$E_{m}= m\Delta_0^2/\mu,\; \,m=0,\pm1,\pm2,...$ \citep{Caroli1964,Volovik1999,Khaymovich2009,Kong2019} where $\Delta_0$ is the pairing far away from any vortices and $\mu$ is the chemical potential. The detection of energetically isolated MVMs at $E_0=0$ requires $E_1=\Delta_0^2/\mu$ to exceed the experimental energy resolution.

\begin{figure}
\noindent \begin{centering}
\includegraphics{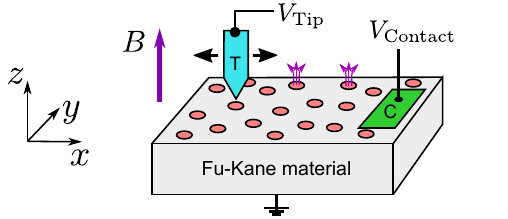}
\par\end{centering}
\caption{\label{fig:Schematic}Schematic of the proposed non-local transport
setup for a Fu-Kane material with superconducting surface Dirac state in the vortex phase. In addition
to the standard scanning-tunneling spectroscopy setup using grounded
bulk and tip contact (``T''), an additional contact (``C'') is
required. This contact does not need to be realized as a second tip but can
be spatially extended.}
\end{figure}

Recently, progress in this direction was made in a variety of novel ``Fu-Kane'' materials that combine bulk superconductivity with two-dimensional surface Dirac states of topological origin and feature $E_1$ on the order of a few hundred $\mu eV$. Prominent example materials with claims for MVM based on LDOS measurements are the iron-based superconductors $\mathrm{FeTe_{0.55}Se_{0.45}}$ \cite{Wang2018,Kong2019,Machida2019,Wang2021a,Zhang2019,Kreisel2020}, $\mathrm{\left(Li_{0.84}Fe_{0.16}\right)}\mathrm{OHFeSe}$\cite{LiuPRX2018, Zhang2021}, $\mathrm{LiFeAs}$\cite{Kong2021}, $\mathrm{CaKFe_4As_4}$\cite{Liu2020} or the transition metal dichalcogenide $\mathrm{2M-WS_2}$\cite{Yuan2019a,Li2021}. 
Besides the agreement of the observed energy spacings with the above theory, another point consistent with the existence of MVMs is the non-oscillatory
radial profile of the MVM-LDOS\citep{Zhang2021}.

On the other hand, for the same sample of $\mathrm{FeTe_{0.55}Se_{0.45}}$, a coexistence of topological vortices and trivial vortices (with CdGM spectra in accordance with $m=\pm\frac{1}{2},\pm\frac{3}{2},...$)
has been reported \citep{Kong2019}. A possible explanation is a high sensitivity of the surface topological superconducting phase
to the exact stoichiometric composition and local chemical potential
\citep{Zhang2019,Li2020,Wang2021a}. There are other concerns regarding the MVM interpretation of experimental results in the putative Fu-Kane materials. They include the possible trivial
origin of a non-split zero-energy vortex bound state \citep{Kim2021},
the sensitivity of the vortex subgap state's energy spacings to the pairing
profile $\Delta(r)$ and to impurities\citep{Chen2021}, or the lack of a robustly
quantized conductance plateau in a strong-coupling transport experiment \citep{Zhu2020}.
%Thus, up to date, the existence of MVMs in the putative Fu-Kane materials remains controversial, and is quite reminiscent of the situation in the aforementioned one-dimensional platforms.

In this work we propose a framework to identify the presence (or absence) of MVMs
in the two-dimensional platform using ideas of non-local transport
first developed for one-dimensional superconducting heterostructures
\citep{Danon2020}. In particular, we propose to use a non-local transport
measurement to spatially map the ratio $[q/n](\mathbf{r})$ of local
charge-density ($q$) and probability-density ($n$) of sub-gap wavefunctions at various
energies. We discuss how the data reveals tell-tale signatures of
either topological MVM or ordinary CdGM states. In contrast to a closely
related pioneering experiment on a one-dimensional quantum wire \citep{Menard2020},
the application of the proposed technique to realistic vortex modes
comes with a number of important modifications: In the one-dimensional wire
case, the spatial resolution is usually limited to the positions of
the tunneling contacts at the two ends of the wire as STM is not applicable. In
the two-dimensional case at least one of the two required surface
contacts can be realized as a movable STM tip (see ``T'' in Fig.~\ref{fig:Schematic}),
which is sufficient to achieve a spatially resolved $q/n$. The second
contact (``C'') can be another STM tip\cite{Li2013-4tip, Clark2013, Clark2014}, if available, or any other
extended type of electrical contact like a patterned metallic overlayer or a graphene flake. 

A second
important difference pertains to the complexity of the electronic
system: While an ideal one-dimensional topological superconductor
harbors two Majorana zero modes at its ends, the two-dimensional situation
is characterized by the fact that vortices (and their putative MVMs)
are located in a disordered lattice with local but essentially random
hybridizations \citep{Cheng2009,Cheng2010} that modify the spectrum from the case of a uniform lattice \citep{Biswas2013,Liu2015b}. Although we start discussing
the most simple case of a single vortex-pair analytically, we then take into account
experimental reality with many vortices using extensive numerical
simulations based on a tight-binding model of the Dirac Hamiltonian.

The rest of the paper is organized as follows: In Sec.~\ref{sec:Model-and-vortex}
we present the low-energy two-dimensional Fu-Kane model and its tight-binding approximation. We then review the description of non-local superconducting quantum
transport in Sec.~\ref{sec:Non-local-transport}. The case of a single
vortex pair is treated in Sec.~\ref{sec:Vortex-pair} which is suitable
to present our protocol proposed for experiments. The applicability
of our main ideas to a realistic disordered vortex lattice is demonstrated
in Sec.~\ref{sec:Distorted-vortex-lattice} and a conclusion is contained
in Sec.~\ref{sec:Conclusion}.

\section{Model and vortex modes\label{sec:Model-and-vortex}}

We consider a single two-dimensional Dirac surface Hamiltonian $\mathcal{H}_{0}=-i\hbar v\left[\sigma_{x}\partial_{x}+\sigma_{y}\partial_{y}\right]-\mu$
with velocity $v$, chemical potential $\mu$ and the $\sigma$-Pauli
matrices acting in spin-space.\footnote{We have assumed for simplicity here that the Dirac point is at the $\Gamma$ point $k=0$ and that the spin-momentum locking is parallel, neither of which is essential.} The second-quantized s-wave pairing
Hamiltonian reads \citep{Fu2008,Cheng2009,Cheng2010}
\begin{equation}
H_{BCS}=\int_{\mathbf{r}}\psi_{\mathbf{r}}^{\dagger}\mathcal{H}_{0}\psi_{\mathbf{r}}+\Delta\psi_{\mathbf{r},\uparrow}^{\dagger}\psi_{\mathbf{r},\downarrow}^{\dagger}+\Delta^{*}\psi_{\mathbf{r},\downarrow}\psi_{\mathbf{r},\uparrow},\label{eq:H_BCS}
\end{equation}
where $\Delta$ is the pairing field and the spinor of electronic
annihilation operators is given by $\psi_{\mathbf{r}}=(\psi_{\uparrow,\mathbf{r}},\psi_{\downarrow,\mathbf{r}})^{\mathrm{T}}$.
The ansatz $\psi_{\mathbf{r},\sigma}\equiv\sum_{n}u_{\sigma,n}\left(\mathbf{r}\right)\gamma_{n}+v_{\sigma,n}^{*}\left(\mathbf{r}\right)\gamma_{n}^{\dagger}$
leads to the following Bogoliubov-de Gennes (BdG) equations for eigenmodes
$\gamma_{n}$ and -energies $E_{n}$,
\begin{align}
\mathcal{H}_{BdG}\Phi(\mathbf{r}) & =E_{n}\Phi(\mathbf{r})\label{eq:BdG}\\
\mathcal{H}_{BdG} & =\tau_{z}(v[\sigma_{x}p_{x}+\sigma_{y}p_{y}]-\mu)+\tau_{x}\mathrm{Re}\Delta-\tau_{y}\mathrm{Im}\Delta
\end{align}
with $\Phi^{\mathrm{T}}(\mathbf{r})=(u_{\uparrow},u_{\downarrow},v_{\downarrow},-v_{\uparrow})$
and Pauli matrices $\tau_{\mu}$ acting in particle-hole space. The
particle-hole symmetry is $\mathcal{P}=\sigma_{y}\tau_{y}\mathcal{K}$
with $\mathcal{P}^{2}=+1$ and $\mathcal{K}$ complex conjugation.
In the homogeneous case, the energies for momentum $\mathbf{k}$ are given by $E_{\mathbf{k}}=\pm(\Delta^{2}+\left(\pm vk-\mu\right)^{2})^{1/2}$. 

A magnetic field $B_{z}$ applied orthogonal to the surface creates
vortices in the pairing field \citep{Chiu2020},
\begin{equation}
\Delta(\mathbf{r})=\Delta_{0}\prod_{j}f(|\mathbf{r}-\mathbf{R}_{j}|)\frac{(x-x_{j})+i(y-y_{j})}{|\mathbf{r}-\mathbf{R}_{j}|}\label{eq:Delta(r)}
\end{equation}
with $\mathbf{R}_{j}=x_j \mathbf{e_x}+y_j \mathbf{e_y}$ the vortex positions and the function $f(r)=\mathrm{tanh}(r/\xi)$
modeling the decay of the pairing amplitude from its bulk value $\Delta_{0}$
towards the vortex core within lengthscale $\xi$. For a single vortex
at the origin, Eq.~\eqref{eq:Delta(r)} reduces to the simple polar-coordinate expression
$\Delta\left(r,\phi\right)=\Delta_{0}f(r)e^{i\phi}$. The magnetic
field can be found from the solution of the London equation which, for
the single vortex case, reads $B_{z}(r)=\frac{\Phi_{0}}{2\pi\lambda^{2}}K_{0}(r/\lambda)$
with corresponding vector potential $\mathbf{A}(\mathbf{r})=\mathbf{e}_{\phi}\frac{\Phi_{0}}{2\pi r}\left[1-\frac{r}{\lambda}K_{1}(r/\lambda)\right]$
in the London gauge. Here, $\Phi_{0}=\pi\hbar/e$ is the magnetic
flux quantum piercing the vortex while the radial decay of $B_{z}(r)$
is controlled by the London penetration depth $\lambda$. The modified
Bessel function of the second kind is denoted by $K_{l}(x)$. The
vector potential enters in the Hamiltonian via the replacement $\mathbf{p}\rightarrow\mathbf{p}-\tau_{z}e\mathbf{A}(\mathbf{r})$.
The generalization to the vector potential for multiple vortices corresponding
to Eq.~\eqref{eq:Delta(r)} is straightforward, $\mathbf{A}(\mathbf{r})\rightarrow\sum_{j}\text{\ensuremath{\mathbf{A}}}(\mathbf{r}-\mathbf{R}_{j})$.

For numerical simulations, we regularize the continuum model on a
two-dimensional square lattice. We set the lattice constant $a=1$,
along with the choice $v=1$, $\hbar=1$. The straightforward regularization
$\mathcal{H}_{0}\rightarrow\mathcal{H}_{0,L}=\sum_{\mathbf{k}}\sigma_{x}\sin k_{x}+\sigma_{y}\sin k_{y}+\sigma_{z}\left(-2+\cos k_{x}+\cos k_{y}\right)-\mu$
can be improved upon replacing $\sin(k)\rightarrow\frac{4}{3}\sin(k)-\frac{1}{6}\sin(2k)$
and $\cos(k)\rightarrow\frac{4}{3}\cos(k)-\frac{1}{3}\cos(2k)$ which
more faithfully approximates the continuum model $\mathcal{H}_{0}$
around $\mathbf{k}=0$ by canceling series expansion coefficients
of order $k_{x}^{3}$ and $k_{y}^{4}$ at the cost of involving hoppings
along bonds $2a\mathbf{e}_{x,y}$. This will ultimately allow us to
choose a large chemical potential ($\mu=0.6$) for the simulations
in the lattice model while still approximating the dispersion of the
continuum model at the Fermi level to a satisfactory degree. This
in turn yields a small length scale for the Fermi wavelength $k_{F}^{-1}$
($\mu=\hbar vk_{F}$) allowing for tractable overall system sizes.
In real space, the lattice Hamiltonian reads
\begin{align}
H_{0,L} & =\sum_{\mathbf{r}}c_{\mathbf{r}}^{\dagger}\left[-2\sigma_{z}-\mu\right]c_{\mathbf{r}}\label{eq:H0L}\\
 & +c_{\mathbf{r}+\mathbf{e_{x}}}^{\dagger}\left[\frac{4}{3}\times\frac{\sigma_{z}+i\sigma_{x}}{2}\right]c_{\mathbf{r}}\nonumber \\
 & +c_{\mathbf{r}+\mathbf{e_{y}}}^{\dagger}\left[\frac{4}{3}\times\frac{\sigma_{z}+i\sigma_{y}}{2}\right]c_{\mathbf{r}}\nonumber \\
 & +c_{\mathbf{r}+2\mathbf{e_{x}}}^{\dagger}\left[-\frac{1}{6}\times\frac{2\sigma_{z}+i\sigma_{x}}{2}\right]c_{\mathbf{r}}\nonumber \\
 & +c_{\mathbf{r}+2\mathbf{e_{y}}}^{\dagger}\left[-\frac{1}{6}\times\frac{2\sigma_{z}+i\sigma_{y}}{2}\right]c_{\mathbf{r}}+h.c.,\nonumber 
\end{align}
and the BdG Hamiltonian becomes 
\begin{equation}
\mathcal{H}_{BdG,L}=\left(\begin{array}{cc}
\mathcal{H}_{0,L} & \Delta\\
\Delta^{*} & -\sigma_{y}\mathcal{H}_{0,L}^{*}\sigma_{y}
\end{array}\right)\label{eq:H_BdGL}
\end{equation}
The inclusion of magnetic field and vortices in the lattice model
is achieved via a discretized version of Eq.~\eqref{eq:Delta(r)}
and the Peierls substitution for the hopping matrix element from $\mathbf{r}_{1}$
to $\mathbf{r}_{2}$ in $H_{0,L}$, 
\begin{equation}
t_{\mathbf{r}_{2},\mathbf{r}_{1}}\rightarrow t_{\mathbf{r}_{2},\mathbf{r}_{1}}\exp\left(\frac{ie}{\hbar}\int_{\mathbf{r}_{1}}^{\mathbf{r}_{2}}d\mathbf{r}\cdot\mathbf{A}(\mathbf{r})\right).\label{eq:Peierls}
\end{equation}
In the limit $\lambda\gg a$, the argument of the exponent can be
approximated by $i\sum_{j}\frac{\theta_{j}(\mathbf{r}_{12})}{2}[1-\frac{r_{j}(\mathbf{r}_{12})}{\lambda}K_{1}(r_{j}(\mathbf{r}_{12})/\lambda)]$
where $r_{j}(\mathbf{r}_{12})\equiv|\mathbf{R}_{j}-(\mathbf{r}_{1}+\mathbf{r}_{2})/2|$
is the distance between the vortex $j$ and the midpoint of the bond
from $\mathbf{r}_{1}$ to $\mathbf{r}_{2}$ and $\theta_{j}(\mathbf{r}_{1,2})$
is the angle between the connection lines $\mathbf{r}_{1,2}-\mathbf{R}_{j}$
measured at the vortex position \citep{Chiu2020}. 

\begin{table}
\noindent \begin{centering}
\begin{tabular}{|c|c|c|c|c|}
\hline 
 & $\hbar v$ & $\mu$ & $\Delta_{0}$ & $k_{F}=\mu/\hbar v$\tabularnewline
\hline 
lat. model & $1$ & $0.6$ & $0.2$ & $1.66$\tabularnewline
\hline 
$\mathrm{FeTe_{x}Se_{1-x}}$ & $25meV\!\cdot\! nm$ & $5meV$ & $1.8meV$ & $0.2/nm$\tabularnewline
\hline 
\hline 
 & $\xi$ & $\zeta=\hbar v/\Delta_{0}$ & $\lambda$ & \tabularnewline
\hline 
lat. model & $2$ & $5$ & $30$ & \tabularnewline
\hline 
$\mathrm{FeTe_{x}Se_{1-x}}$ & $4.6nm$ & $13.9nm$ & $500nm$ & \tabularnewline
\hline 
\end{tabular}
\par\end{centering}
\caption{\label{tab:params} Summary of parameters used for the lattice model simulations and for the experimentally realized material $\mathrm{FeTe_{x}Se_{1-x}}$,
$ x \simeq 0.55$ as compiled in Ref.~\onlinecite{Chiu2020}. Here, $v$ and $\mu$ are the velocity and chemical potential of the Dirac surface Hamiltonian, respectively. For the lattice model, we set $\hbar v=1$
and $a=1$ for the lattice constant. The surface state pairing amplitude without vortices is given by $\Delta_0$ while $\xi$ denotes the length-scale on which the pairing decays towards vortex cores. The superconducting coherence length is $\zeta$ and the London penetration length is denoted by $\lambda$. %\mb{How do the experimental values compare to the lattice model? Can they be put in the same units?} 
}
\end{table}

The MVM wavefunction for a single vortex in the continuum model reads \citep{Cheng2009,Cheng2010}
\begin{equation}
\Psi(r,\phi)\propto\exp\left[-\zeta^{-1}\int_{0}^{r}\mathrm{d}p\,f(p)\right]\left(\begin{array}{c}
e^{-i\pi/4}J_{0}\left(rk_{F}\right)\\
e^{+i\pi/4+i\phi}J_{1}\left(rk_{F}\right)\\
e^{-i\pi/4-i\phi}J_{1}\left(rk_{F}\right)\\
-e^{+i\pi/4}J_{0}\left(rk_{F}\right)
\end{array}\right)\label{eq:MVM}
\end{equation}
where $J_{l}(x)$ is the Bessel function of the first kind and the decay in radial direction is governed by the
Majorana coherence length is $\zeta=\hbar v/\Delta_{0}$. Here, the
effect of the vector potential $\mathbf{A}(\mathbf{r})$ has been
neglected as justified for a single vortex if $\lambda\gg\zeta$.

\begin{figure}
\begin{centering}
\includegraphics{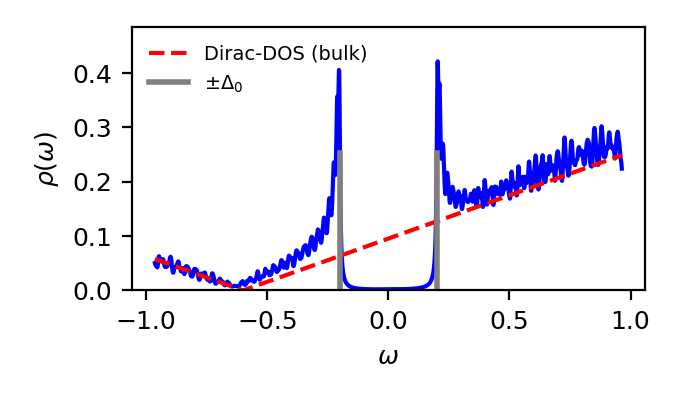}
\par\end{centering}
\caption{\label{fig:bulkLDOS}Tunneling LDOS $\rho(\omega)=\int d\mathbf{r}\rho(\omega,\mathbf{r})/\int d\mathbf{r}$
for the two-dimensional Fu-Kane model in the absence of magnetic
field as found from exact diagonalization of the lattice model \eqref{eq:H_BdGL}
with parameters $\mu=0.6$ and $\Delta_{0}=0.2$ and system size $L_{x}\times L_{y}=84\times86$.
The oscillations are due to finite-size effects which are incompletely
smoothed by the intrinsic level broadening $\Gamma_{0}=0.02\Delta_{0}=0.004$.
The superconducting gap $\omega=\pm\Delta_{0}$ is indicated by grey
vertical lines and the density of states (DOS) of the normal-state
Dirac Hamiltonian $D(\omega)=\frac{\omega}{2\pi(\hbar v)^{2}}$ is
depicted by the red dashed line.}
\end{figure}

We choose 
the lattice model parameters as $\mu=0.6$, $\Delta_{0}=0.2$, $\xi=2$, $\lambda=30$, the unit of energy is given by $\hbar v/a=1$
and the unit of length is $a=1$. As summarized in Tab.~\ref{tab:params}, this choice of parameters is motivated by comparison to the experimentally extracted values for $\mathrm{FeTe_x Se_{1-x}}$, which are of similar relative size. Only the London penetration length $\lambda$ of the lattice model, while still being by far the largest length scale, is chosen smaller than what would be appropriate in $\mathrm{FeTe_x Se_{1-x}}$ to keep the required lattice sizes tractable. The one-dimensional gapless Majorana
mode localized at the open boundaries of the system does not affect the results
below due to sufficient distance between vortices and boundary, so that the hybridization
between vortex bound states and the edge modes is negligible compared
to inter-vortex hybridizations. The LDOS $\rho(\omega)$ (see Eq.~\eqref{eq:rho} below
for a definition) of the finite-size lattice
model without vortices and averaged in the center region  is shown in Fig.~\ref{fig:bulkLDOS} and agrees
to the expectation from the continuum model. Further, we have checked
that the wavefunction obtained numerically for a single vortex zero-mode
agrees with the analytic prediction for the MVM in Eq.~\eqref{eq:MVM}
and that the first excited CdGM-state appears at an energy of order
$0.09\sim\Delta^{2}/\mu$ as predicted by theory.\cite{Caroli1964,Khaymovich2009}

\section{Non-local transport\label{sec:Non-local-transport}}

We now consider a transport setup and attach an STM tip ``T'' as well as
a ground contact, see Fig.~\ref{fig:Schematic}
(contact ``C'' is to be added at a later stage, see below). For concreteness and to
set the stage for the lattice model simulations using the \noun{kwant}
software package\citep{Groth2014}, we model the tip ``T'' as a
one-dimensional chain of single sites with hopping $t=1$ diagonal
in spin space. This choice will provide a density of states that does not vary appreciably over the small range of bias $|\omega|\ll1$ applied in the following. The lead is locally coupled to the surface with hopping $\gamma_{T}$ which reflects the tip-sample tunneling matrix element \cite{Tersoff1983} but will be chosen in an ad-hoc manner below as we are not aiming to model a specific setup.
The tip-induced broadening of an eigenstate $\Phi$ is $\Gamma_{T}\equiv\gamma_T^{2}n_T/t$
where $n_{T}\equiv\sum_{\sigma}|u_{\sigma}(\mathbf{r}_{T})|^{2}+|v_{\sigma}(\mathbf{r}_{T})|^{2}$
is the eigenstate intensity at the lead position.

With the exception of the strong-tunneling experiment by Zhu\emph{
et al}.\citep{Zhu2020}, all existing experimental or numerical transport
studies of the Fu-Kane setup were done at weak
coupling to the tip. This regime is characterized by a tip-induced
broadening $\Gamma_{T}$ which is smaller than the intrinsic
relaxation rate $\Gamma_{0}$ of the quasiparticles. This means that
an injected quasiparticle decays in the sample before it can return
to the lead. In the case when the intrinsic broadening exceeds the
thermal broadening from the leads, $T\lesssim\Gamma_{0}$ ($k_{B}=1$),
we obtain the broadened LDOS from\citep{Martin2014}
\begin{eqnarray}
\rho\left(\omega,\mathbf{r}\right) & = & \sum_{E_{n}>0}\frac{\Gamma_{0}/\pi}{\left(\omega-E_{n}\right)^{2}+\Gamma_{0}^{2}}\sum_{\sigma=\uparrow,\downarrow}\left|u_{n,\sigma}(\mathbf{r})\right|^{2}\label{eq:rho}\\
 & + & \sum_{E_{n}>0}\frac{\Gamma_{0}/\pi}{\left(\omega+E_{n}\right)^{2}+\Gamma_{0}^{2}}\sum_{\sigma=\uparrow,\downarrow}\left|v_{n,\sigma}(\mathbf{r})\right|^{2}.\nonumber 
\end{eqnarray}
where we choose $\Gamma_{0}=0.02\Delta_{0}=0.004$ in Fig.~\ref{fig:bulkLDOS}
and for the simulations below. The LDOS is proportional to the differential
conductance $dI/dV$ at the bias $\omega=eV$ relative to the ground
contact, see Fig.~\ref{fig:Schematic}. In light of Eq.~\eqref{eq:rho},
the LDOS yields information about the eigenenergies of the system
and the spatial distribution of their wavefunction's electron and
hole content. In particular, it cannot distinguish a MVM at $E_{0}=0$
from an ordinary excitation with energy $E_{n}>0$ but smaller than
$\Gamma_{0}$ or $T$.

The experiment of Zhu\emph{ et al}.\citep{Zhu2020} reached the strong
coupling regime $\Gamma_{T}>\Gamma_{0}$ where quasiparticle transport
becomes (approximately) coherent and can be described by a unitary
scattering matrix formalism.\citep{NazarovBlanterBook} Due to the
bulk superconducting gap, the quasiparticles at subgap energies solely
enter and leave through the tip. In the presence of a MVM, perfect
Andreev reflection is expected at zero bias which, according to theory\citep{Law2009,Flensberg2010},
should yield $G\equiv dI/dV=2e^{2}/h$. As this result should be independent
of details, a plateau in $G$ as a function of tip-sample separation
is expected. It is currently an open question why the experimental
conductance plateaus\citep{Zhu2020} typically show a significantly
smaller value for $G$ that varies from vortex to vortex. 

We now describe the three-terminal transport setup analyzed in the
remainder of this work. We add a second lead (``contact C'') at
the sample surface in the vicinity of the tip ``T'', see Fig.~\ref{fig:Schematic}.
We keep the assumption of strong coupling, $\Gamma_{T,C}>\Gamma_{0}$. At
subgap energies $|\omega|\leq\Delta_{0}$, this opens up a multitude
of quasiparticle scattering channels where electrons and holes can
enter or leave via either lead, provided there is an eigenstate of the isolated sample with
simultaneous support at both lead positions. The objects of interest
are the (dimensionless) conductances $g_{\alpha\beta}\equiv dI_{\alpha}/dV_{\beta}/[e^{2}/h]$
where $I_{\alpha}$ is the electrical current flowing into lead $\alpha=\{C,T\}$
and $V_{\beta}$ is the bias at lead $\beta$. The scattering matrix
for this non-local setup mediated by a single eigenstate at arbitrary energy
$E_{0}$ was analyzed by Danon\emph{ et al}\citep{Danon2020} for
the case of spinless electrons. In Appendix \ref{app:Scattering-matrix} we
generalize this analytical calculation to the case with spin, but the
quantitative behavior of the conductances close to the resonance $|\omega|\simeq E_{0}$
is not affected by this modification. Focusing on equal bias voltage
for the two leads, one can approximate the non-local zero-temperature
conductance as\citep{Danon2020}
\begin{equation}
g_{CT}\left(\omega\simeq\pm E_{0}\right)\simeq\frac{-8\xi_{C}E_{0}}{\left[\omega^{2}-E_{0}^{2}\right]^{2}+4\Gamma^{2}E_{0}^{2}}\left(E_{0}\xi_{T}+\omega\Gamma_{T}\right).\label{eq:g_CT}
\end{equation}
Here, $\Gamma\equiv\Gamma_{C}+\Gamma_{T}$ is the sum over the two
lead-induced level broadenings $\Gamma_{\alpha}\equiv\gamma_{\alpha}^{2}n_{\alpha}/t$
where $n_{\alpha}\equiv u_{\alpha}+v_{\alpha}$ is the total wavefunction
intensity at the contact position, with $u_{\alpha}\equiv\sum_{\sigma}|u_{\sigma}(\mathbf{r}_{\alpha})|^{2}$
and $v_{\alpha}\equiv\sum_{\sigma}|v_{\sigma}(\mathbf{r}_{\alpha})|^{2}$.
It is assumed that $\Gamma\ll E_{0}$ for Eq.~\eqref{eq:g_CT} to
hold. The quantity $\xi_{\alpha}\equiv\gamma_{\alpha}^{2}q_{\alpha}/t$
is proportional to the local BCS-charge $q_{\alpha}\equiv u_{\alpha}-v_{\alpha}$
which is of central interest in the following discussion. We emphasize that Eq.~\eqref{eq:g_CT} describes transport mediated by an extended state in the superconducting gap ($E_0<\Delta$) where transport through the superconducting bulk is suppressed.

The crucial observation in Eq.~\eqref{eq:g_CT} is the asymmetry
of the two peak heights $\omega\simeq\pm E_{0}$ due to the second
term in parenthesis which is odd in $\omega$. We define the symmetric
and asymmetric part of the non-local conductance as $g_{CT}^{\mathrm{sym/asym}}\left(\omega\right)\equiv\frac{1}{2}\left[g_{CT}\left(\omega\right)\pm g_{CT}\left(-\omega\right)\right]$
and observe\citep{Danon2020} from Eq.~\eqref{eq:g_CT}
\begin{equation}
\frac{g_{CT}^{\mathrm{sym}}}{g_{CT}^{\mathrm{asym}}}(\omega\simeq E_{0})\simeq\frac{q_{T}}{n_{T}}=\frac{|u_{T}|^{2}-|v_{T}|^{2}}{|u_{T}|^{2}+|v_{T}|^{2}}\in[-1,1].\label{eq:q/n}
\end{equation}
This relation allows for the extraction of $q_{T}/n_{T}$, the ratio
of BCS-charge and intensity of an eigenstate at energy $E_{0}$ at
the tip position. The prerequisite is that a pair of peaks at $\omega\simeq\pm E_{0}$
can be identified in the $g_{CT}$ data. 

The significance of the quantity $q_{T}/n_{T}$ for detecting Majorana
zero modes lies in the fact that an isolated Majorana zero mode fulfills
$q_{T}(\mathbf{r})=0$ at every position $\mathbf{r}$ due to the particle-hole symmetric nature
of the state. On the other hand, for an isolated Majorana zero mode
at $E_{0}=0$, the condition $\Gamma\ll E_{0}$ cannot be achieved.
Consequently, one has to rely on the hybridization between zero modes
to push the energy $E_{0}$ to finite values so that $q_{T}/n_{T}$
can be detected by non-local transport, thereby compromising $q_{T}=0$ to a certain degree.
In the following, we apply these general ideas to the case of MVMs
and show that MVMs set themselves apart from the CdGM-states at finite
energy by a peculiar spatial signature of the $q_{T}/n_{T}$ map.

Based on the above discussion and Eq.~\eqref{eq:g_CT} we discuss the requirements for the
second contact ``C''. While the achievable spatial resolution of $q_T/n_T$ hinges on the sharpness and movability offered by the STM-tip ``T'', the contact ``C'' can be stationary. In particular,
if no multiple-tip STM instrument is available\cite{Li2013-4tip}, the contact can even
be spatially extended. In light of Eq.~\eqref{eq:g_CT}, such an extended contact will reduce the risk of hitting a contact position where $\xi_C \sim q_C \simeq 0$ which would cause a vanishing non-local transport signal. On the other hand, as we require $\Gamma=\Gamma_T+\Gamma_C \ll E_0$ for Eq.~\eqref{eq:g_CT} to hold in the first place, we must limit the contact-induced level broadening which grows with contact area and density of states. It might thus be beneficial to choose a contact material with a low density of states, like a graphene flake, or limit the size of the contact using nanofabrication techniques. For example, local gold nanocontacts can be made at selective surface sites using STM via a field-induced atomic emission process in situ.\cite{Qin2012,Qin2012a} A discussion on the role of the contact-tip distance is postponed to the end of Sec.~\ref{sec:Distorted-vortex-lattice}.

We now turn to the leading effect of temperature on the quasiparticle structure, assuming that the temperature remains low enough that the superconducting properties and vortex locations are unmodified. First, the sample temperature $T_s$ needs to be small enough so that the temperature dependent intrinsic quasiparticle decay $\Gamma_{0}$
can be neglected against $\Gamma_{T,C}$ for our coherent non-local transport theory to apply. We next consider the effective electron temperature $T_{\text{eff},\alpha}$ in lead $\alpha$ which usually exceeds the sample temperature (Ref.~\onlinecite{Machida2018} determined 85mK for the former and about 40mK for the latter in the case of an STM tip). Theoretically, $T_{\text{eff},\alpha}$ is taken into account by a convolution of $g_{\alpha\beta}(\omega)$ with the derivative of the Fermi function $-\frac{df(\omega,T_{\text{eff},\beta})}{d\omega}=\frac{1}{4T_{\text{eff},\beta}}\cosh^{-2}\left(\frac{\omega}{2T_{\text{eff},\beta}}\right)$. Since the $V_\beta$-dependence of $I_\alpha$ is assumed to only enter via the distribution functions of the leads,\cite{Danon2020} the broadening procedure of $g_{CT}$ in Eq.~\eqref{eq:g_CT} is to be applied with the effective electron temperature of the tip ``T'', $T_{\text{eff},T}$. The latter will be abbreviated simply as ``temperature'' $T$ in the following.    

At zero temperature, the peaks of Eq.~\eqref{eq:g_CT} which occur at $\omega = \pm E_{0}$
have the same width $2\Gamma$. Hence their temperature broadened amplitudes
are diminished simultaneously for both signs of $\omega$. If temperature
reaches the scale $E_{0}$, the broadening symmetrizes the overall
trace $g_{CT}\left(\omega\right)$ leading to a underestimation of
$|g_{CT}^{\mathrm{asym}}|$ as compared to its $T=0$ value. As
a consequence, the quantity $|g_{CT}^{\mathrm{sym}}/g_{CT}^{\mathrm{asym}}(\omega\simeq E_{0})|$
can then exceed unity in magnitude which should be taken as a warning
that the right-hand side of Eq.~\eqref{eq:q/n} no longer applies.

In the following we theoretically implement the above protocol. We
assume that $\Gamma_{0}$ is sufficiently small so that the scattering
matrix approach is justified. However, we take into account a finite
temperature in the leads. While the case with two vortices studied
in the subsequent Sec.~\ref{sec:Vortex-pair} is still analytically
tractable, our numerical approach is particularly useful for the realistic
case of a distorted vortex lattice. Here the above assumption of a
single spectrally isolated subgap state at energy $E_{0}$ drastically
fails as every pair of MVMs contributes one fermionic state that cluster
in a MVM- or CdGM-band. However, our exact numerics still shows that
the peculiar signatures found for the vortex-pair still survive in the vortex-lattice $g_{CT}^{\mathrm{sym}}/g_{CT}^{\mathrm{asym}}$
map.

\section{Vortex pair\label{sec:Vortex-pair}}

\begin{figure*}
\noindent \begin{centering}
\includegraphics{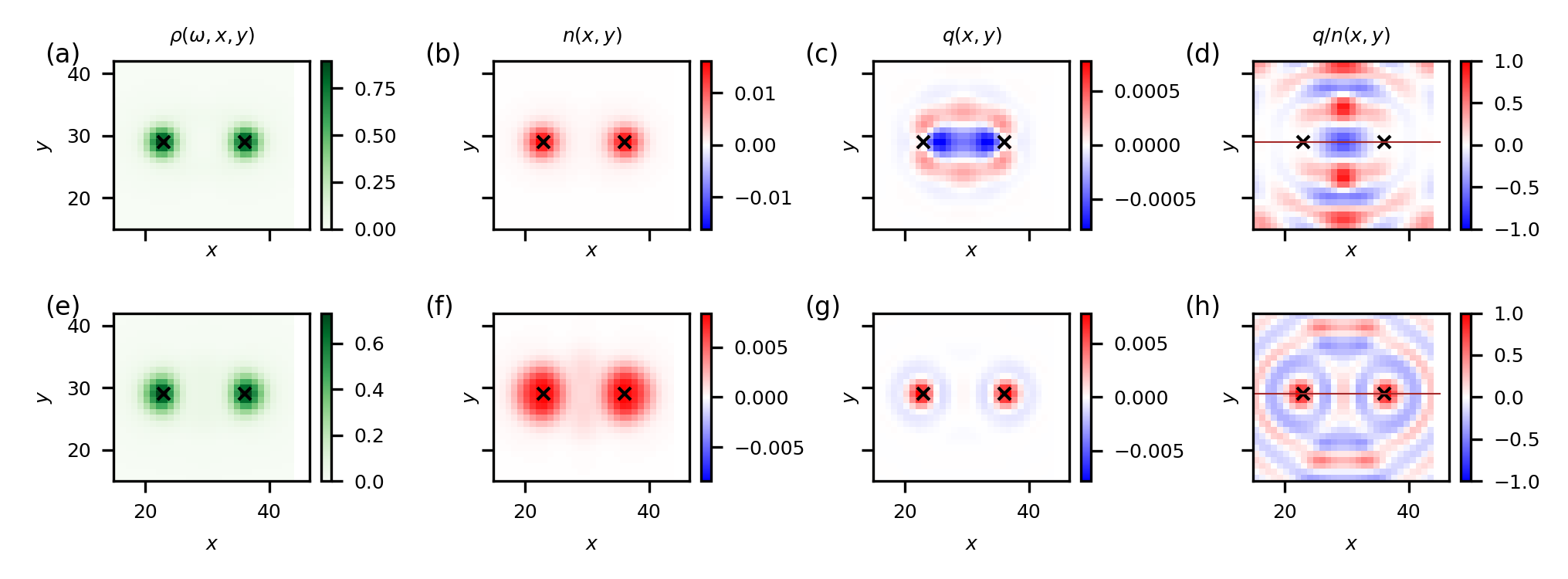}
\par\end{centering}
\caption{\label{fig:pair_ED}Numerical lattice-model results from exact diagonalization
for a pair of vortices located at a distance $R=13$ as indicated
by the black crosses. The parameters are $\mu=0.6$, $\Delta_{0}=0.2$,
$\xi=2$ and $\lambda=30$ and the overall system size is $L_{x}\times L_{y}=60\times58$.
The top row shows the LDOS $\rho$ {[}Eq.~\eqref{eq:rho} with $\Gamma_{0}=0.004$,
$\omega=E_{0}${]} in panel (a), intensity $n$ in panel (b), charge $q$ in panel (c) and the ratio $q/n$ in panel (d)
for the hybridized MVM state at energy $E_{0}=0.0045$, the bottom
row with panels (e-h) reports the same quantities for the lower one of the two hybridized
CdGM-state energies, $E_{1}=0.09$.}
\end{figure*}
\begin{figure}
\noindent \begin{centering}
\includegraphics{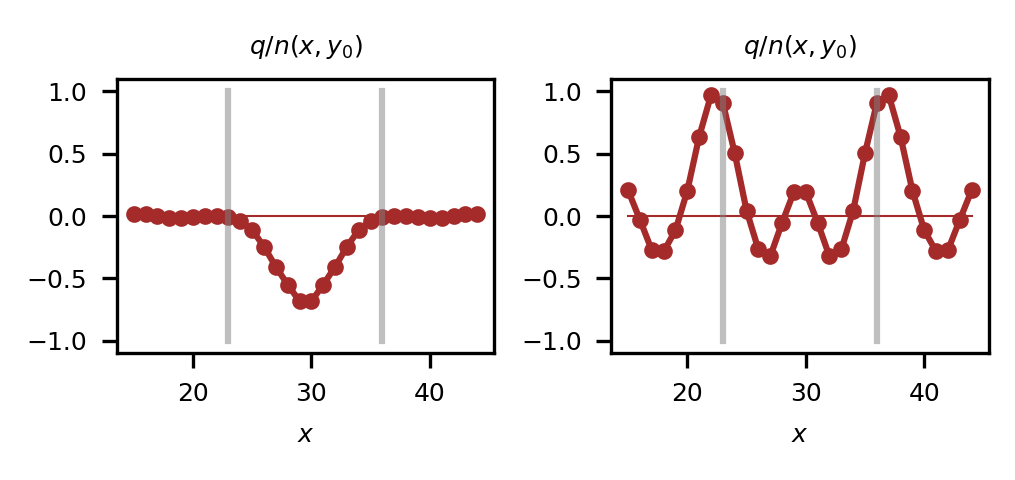}
\par\end{centering}
\caption{\label{fig:pair_ED-cut}One-dimensional cut through the data depicted
in Fig.~\ref{fig:pair_ED}(d,h) along the brown line connecting
the two vortices. The vertical lines denote the vortex positions.
The low energy MVM data from Fig.~\ref{fig:pair_ED}(d) is shown in the left panel, the right panel displays
the hybridized finite energy CdGM-state from Fig.~\ref{fig:pair_ED}(h).}
\end{figure}

We now investigate the case of a single pair of vortices where for
the hybridized MVMs, we can find $[q/n](\mathbf{r})$ analytically from
the single MVM wavefunction, Eq.~\eqref{eq:MVM}. We place the vortices
at positions $\mathbf{R}_{1,2}=\mathbf{R}_{0}\pm R/2\mathbf{e}_{x}$
and use two sets of polar coordinates $r_{j}=|\mathbf{r}-\mathbf{R}_{j}|$
and $\phi_{j}=\mathrm{arg}\left(\mathbf{r}-\mathbf{R}_{j}\right)$
for $j=1,2$. The hybridized MVM states\citep{Cheng2010} can be approximated
by $\Psi_{s=\pm}=(\Psi_{1}+si\Psi_{2})/\sqrt{2}$ where the phase
of the pairing field $\Delta$ just left to each vortex is $\Omega_{1}=0$
and $\Omega_{2}=\pi$, which is taken into account by a relative prefactor $e^{i\tau_{z}\Omega_{j}/2}$
between $\Psi_{j}$ and Eq.~\eqref{eq:MVM}. Dropping the wavefunction
normalization, we obtain for the profile of the intensity and charge density
\begin{eqnarray}
n_{s}(\mathbf{r}) & \propto & e^{-2r_{1}/\zeta}\left[J_{0}^{2}(r_{1}k_{F})+J_{1}^{2}(r_{1}k_{F})\right]+r_{1}\!\rightarrow\!r_{2},\label{eq:n(r)_pair}\\
q_{s}(\mathbf{r}) & \propto & -2se^{-\left(r_{1}+r_{2}\right)/\zeta}\{J_{0}(r_{2}k_{F})J_{0}(r_{1}k_{F})\label{eq:q(r)_pair}\\
 &  & +\cos(\phi_{1}-\phi_{2})J_{1}(r_{1}k_{F})J_{1}(r_{2}k_{F})\}.\nonumber 
\end{eqnarray}
Note that $n_{s}(\mathbf{r})$ is proportional to sum of the two individual MVM's intensities, qualitatively similar to the LDOS $\rho(\omega,\mathbf{r})$,
see Fig.~\ref{fig:pair_ED}(a,b), for the corresponding plots
based on exact diagonalization (ED) of the lattice model. In
contrast, the spatial structure of $q_{s}(\mathbf{r})$ is dominated
by the exponential prefactor which gives rise to an ellipsoidal structure
with the two vortices in the focal points and oscillations caused
by the remaining terms. 

We expand the Bessel functions at a sufficient distance from the vortices
$r_{1,2}k_{F}\gg1$. We further restrict to a point $\mathbf{r}_{c}$
on the connecting line between the vortices, where $\phi_{1}=\pi$,
$\phi_{2}=0$, $r_{1}+r_{2}=R$ and obtain
\begin{equation}
\frac{q_{s}}{n_{s}}(\mathbf{r}_{c})=\frac{-2se^{-R/\zeta}\frac{1}{\sqrt{r_{1}r_{2}}}\sin(k_{F}R)}{\frac{1}{r_{1}}\exp\left(-2r_{1}/\zeta\right)+\frac{1}{r_{2}}\exp\left(-2r_{2}/\zeta\right)}\label{eq:q/n(r_c)}
\end{equation}
which is peaked at the mid-point $\bar{\mathbf{r}}_{c}=(\mathbf{R}_{1}+\mathbf{R}_{2})/2$.
The peak value is $q_{s}/n_{s}(\bar{\mathbf{r}}_{c})=-s\sin(k_{F}R)$
which oscillates like the MVM hybridization\citep{Cheng2010} $E_{0}\sim\cos\left[k_{F}R+\frac{1}{2}\tan^{-1}(\zeta k_{F})\right]$
with a relative phase shift depending on the value of $\zeta k_{F}$ and valid for $R \gg \xi,1/k_F $. Note that for the theoretically interesting case of $\mu = 0$, which is unrealistic in current materials, chiral symmetry prevents hybridization
($E_0 = 0$) for vortices of the same vorticity.\citep{Cheng2010}
For the lattice model with vortex distance $R=13$ and $\mu=0.6$, we present $q_{s}(\mathbf{r})$
and $[q_{s}/n_{s}](\mathbf{r})$ of the hybridized MVM state with $E_{0}=0.0045$
in Fig.~\ref{fig:pair_ED}(c,d). The data for $q_{s}/n_{s}$
on the cut between the two vortices is depicted in Fig.~\ref{fig:pair_ED-cut} (left)
and shows good qualitative agreement with the analytical prediction
above. A quantitative comparison is complicated due to an inaccuracy
of the ansatz $\Psi_{s=\pm}=(\Psi_{1}+si\Psi_{2})/\sqrt{2}$ as documented
by a slight renormalization of the wavefunction peak-intensity separation
beyond the vortex distance $R$ (data not shown). We remark that in one-dimensional proximitized semiconductor quantum wires a pair of hybridized Majorana bound states is expected to cause a qualitatively similar form for the fraction $\frac{q_s}{n_s}(x)$   \cite{Ben-Shach2015, Danon2020}.

We now discuss the numerical lattice-model ED results for $\rho,n,q$
and $q/n$ as obtained for one of the two hybridized first excited
states of each vortex which are split around $E\simeq0.1$, see Fig.~\ref{fig:pair_ED}(e-h) and Fig.~\ref{fig:pair_ED-cut} (right). While the spatial structure
of $\rho$ and $n$ are qualitatively indistinguishable from the MVM
case, $q(\mathbf{r})$ shows local maxima around the
two vortex positions with radially oscillating signs. This resembles
the sum of $q(\mathbf{r})$ of the solutions individual to each vortex.
Note that the structures of $q/n$ at larger distances from the vortices
shown in Fig.~\ref{fig:pair_ED}(d,h) emerge from the ratio
of two numbers very small in magnitude and are likely unobservable
in a non-local transport experiment due to insufficient peak visibility and intrinsic broadening,
c.f.~Eq.~\eqref{eq:g_CT}.

In summary, based on the elementary case of a vortex pair, we propose
to identify hybridized states of MVMs by their non-local spatial distribution
of $q/n$ which attains values close to zero at the vortex positions
and magnitudes attaining their maxima in between. In contrast, ordinary
CdGM-states show peaks of $|q/n|$ at the vortex positions. The positions
of the vortices can be experimentally obtained from the LDOS $\rho(\omega,\mathbf{r})$
map as usual,\citep{Machida2019} while the information on $q/n$
can be obtained experimentally from the non-local transport measurement via $g_{CT}^{\mathrm{sym}}/g_{CT}^{\mathrm{asym}}$ at an energy $\omega=E_0$ where $g_{CT}$ peaks. While this relation could be shown analytically for the case of a single energetically well-separated subgap state (i.e.~the vortex-pair case), it remains valid qualitatively for the case of a band of subgap states as in the case of a distorted vortex lattice as we show below.

In the remainder of this paper, we will demonstrate the above assertion using a numerical implementation of the non-local transport measurement on a
faithful lattice model with finite-temperature leads attached. 
We start with the vortex pair, see Fig.~\ref{fig:pair_transport}.
The contact ``C'' is placed at the top boundary of the
field of view, in the vicinity of the vortices (green patch). In panel (d)
we show the resulting $g_{CT}(\omega,\mathbf{R}_{1})$ at
the position of the right vortex, panel (a) zooms into small energies.
The non-local conductance shows temperature broadened peaks and dips
at $|\omega|=E_{0}=0.0045$ and $|\omega|=E_{1}=0.09$
indicating the energies of the hybridized MVM- and CdGM-states in
agreement with the ED results [see dashed vertical lines in panels (a) and (d)]. Panels (b) and (e) show a spatial map of $g_{CT}(\omega=E_{0,1},\mathbf{r})$,
respectively. The ratio $g_{CT}^{\mathrm{sym}}/g_{CT}^{\mathrm{asym}}(\omega=E_{0,1})$
for both peak positions is shown in  Fig.~\ref{fig:pair_transport} panels (c) and (f), respectively.
The agreement with the ED results in Fig.~\ref{fig:pair_ED} is excellent
in almost the entire field of view, confirming the practical applicability
of Eq.~\eqref{eq:q/n}.

\noindent \begin{center}
\begin{figure*}
\noindent \begin{centering}
\includegraphics{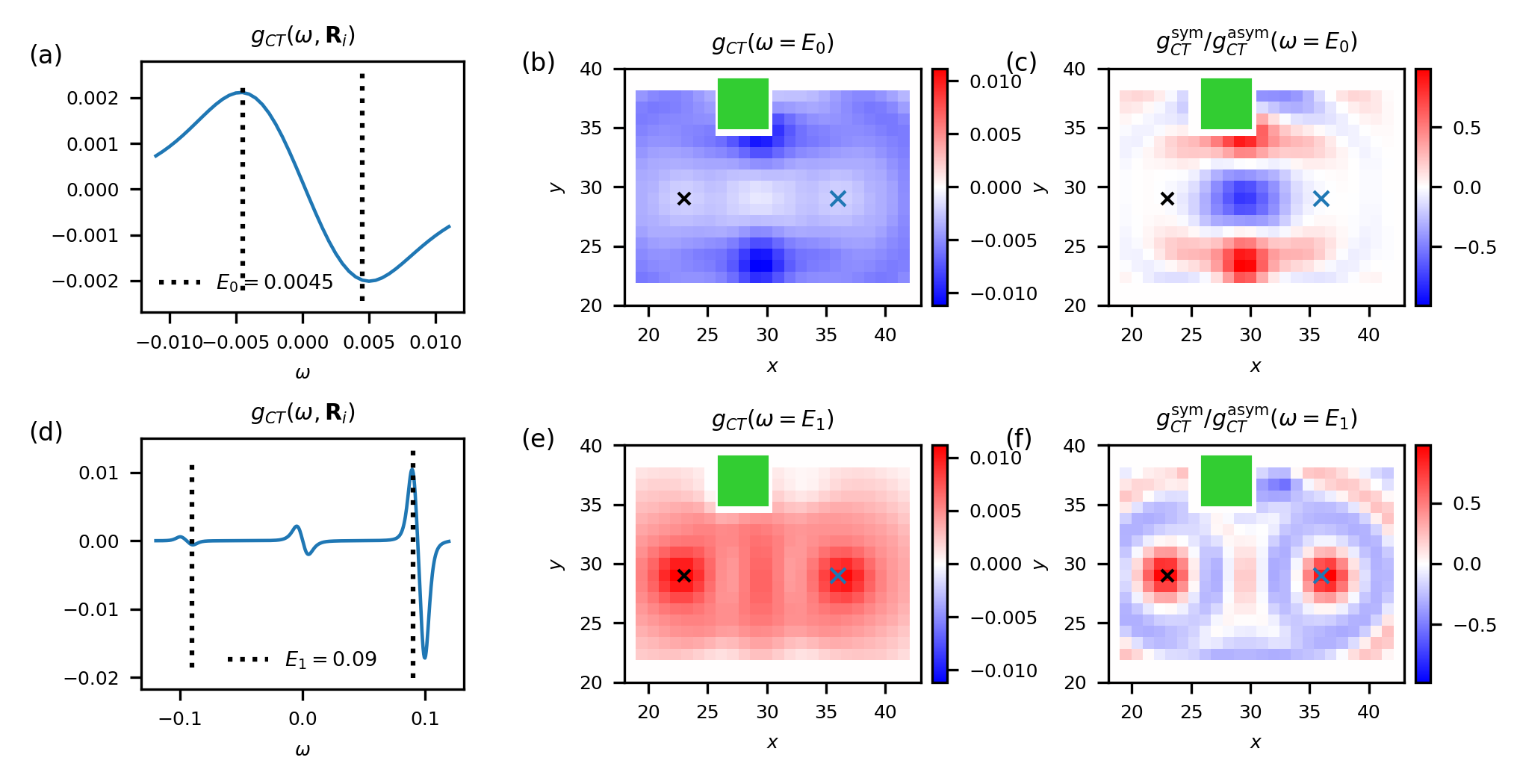}
\par\end{centering}
\caption{\label{fig:pair_transport}Non-local transport simulation for a lattice
model including a vortex pair. We model the tip ``T'' as a single-atomic
lead with $t=1$ and $\gamma_{T}=0.4$ and the extended contact ``C''
as a $A_C=4 \times 4$ patch of the same single-atomic leads with $\gamma_{C}=0.1$
(green). For a local intensity $n_C$ on the order of $0.01$ (c.f. Fig.~\ref{fig:pair_ED})
this results in a broadening $\Gamma_C=n_C A_C \gamma_{C}^{2}/t\simeq0.002$
which is smaller than the MVM hybridization energy (we neglect the
intrinsic broadening $\Gamma_{0}$ in order to obtain a unitary scattering
matrix). For the temperature of the leads, we take $T=0.002$. The
model parameters (see Tab.~\ref{tab:params}) are the same as in Fig.~\ref{fig:pair_ED}. The
left column with panels (a,d) shows the bias-dependent non-local conductance $g_{CT}(\omega,\mathbf{R}_{1})$
with the tip positioned at the right vortex. A low-energy peak structure highlighted by the vertical dashed lines
appears at $|\omega|=E_{0}=0.0045$, see panel (a), and $|\omega|=E_{1}=0.09$, see panel (d). In panels (b) and (e), we depict $g_{CT}(\omega=E_{0,1},\mathbf{r})$.
The right panels (c) and (f) depict $g_{CT}^{\mathrm{sym}}/g_{CT}^{\mathrm{asym}}(\omega=E_{0,1})$
which quantitatively agree to the $q/n$ maps of Fig.~\ref{fig:pair_ED}(d,h).}
\end{figure*}
\par\end{center}

\section{Distorted vortex lattice\label{sec:Distorted-vortex-lattice}}
\noindent \begin{center}
\begin{figure*}
\noindent \begin{centering}
\includegraphics{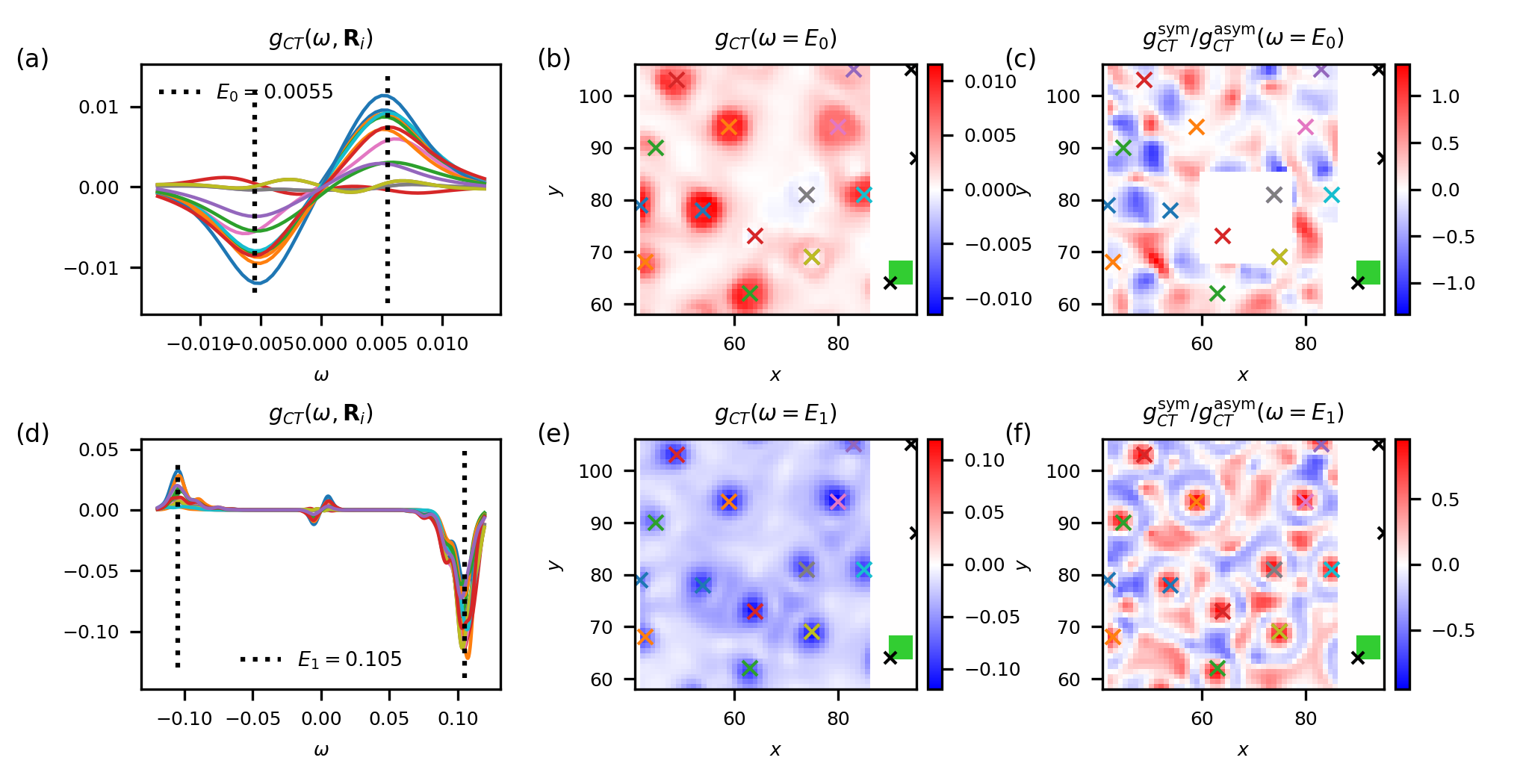}
\par\end{centering}
\caption{\label{fig:vortexLattice_transport}Non-local transport simulation
for a lattice model including a set of vortices arranged in a distorted lattice. The
parameters are the same as in Fig.~\ref{fig:pair_transport}. The left column with panels (a,d) shows the bias-dependent non-local conductance $g_{CT}(\omega,\mathbf{R}_{i})$ for thirteen
vortices $i=1,2,...,13$. The colors of the lines correspond to the color of the crosses in the other panels, the latter mark the vortex positions. Panel (a) zooms to small bias voltages around the peak at $E_0=0.0055$ while panel (d) shows a larger bias range including the peak at $E_1=0.105$. In panels (b) and (e), we depict $g_{CT}(\omega,\mathbf{r})$ for $\omega=E_{0}$ and $\omega=E_{1}$, respectively. The right panels (c) and (f) depict $g_{CT}^{\mathrm{sym}}/g_{CT}^{\mathrm{asym}}$ for $\omega=E_{0}$ and $\omega=E_{1}$, respectively. As the inner three vortices (red, grey, yellow) do
not show a pronounced peak structure at low energies $\sim E_{0}$ in panels (a,b),
we discard their vicinity for the plot in panel (c).}
\end{figure*}
\par\end{center}

We now turn to the experimentally realistic case of a distorted vortex
lattice. Owing to the presence of a finite density of states both
in the MVM- and CdGM-band, the analytical treatment from Sec.~\ref{sec:Non-local-transport}
building on the presence of a single spectrally isolated eigenstate
a priori does not apply any longer and we resort to numerical simulations.
We use a large sample $L_{x}\times L_{y}=160\times162$ with an average
vortex distance $R\sim13$ similar to the separation of the vortex
pair studied above. To avoid edge effects, we focus on the central
region of the sample. In Fig.~\ref{fig:vortexLattice_transport},
the vortex positions in the central region (which could be found experimentally
via the LDOS $\rho$) are denoted by crosses. The contact
(green patch) is placed on the bottom right relative to the scanning-tip
field of view which includes thirteen vortices (colored crosses). The
data for $g_{CT}(\omega,\mathbf{R}_{j})$ at these vortex positions
are shown in panel (d) and panel (a) shows a zoom-in about low energies where the hybridized MVMs occur. We observe a peak structure at $|\omega| \simeq E_{0}=0.0055$
for the outer ten out of the thirteen vortex positions and at $|\omega| \simeq E_{1}=0.105$
for all vortex positions in the field of view. Panels (b) and (e) show
the spatially resolved $g_{CT}(\omega,\mathbf{r})$ for $\omega=E_{0,1}$, respectively.
Our main result is shown in panels (c) and (f). Here we report $g_{CT}^{\mathrm{sym}}/g_{CT}^{\mathrm{asym}}(\omega=E_{0,1})$
which qualitatively resembles the observations made for the vortex
pair: For the MVM band around $\omega=E_{0}$, the data in panel (c) vanishes at and around the ten outer vortex positions and shows extended non-local features in
between vortices. We disregard the region around the three central
vortices for which no sizable peak structure was
observed in the first place. Presumably, the reason for the local absence of sizable peaks
is that the MVM-band does not contain
a state simultaneously supported in the region of the central three vortices and at the
contact ``C''. 

The signatures of $g_{CT}^{\mathrm{sym}}/g_{CT}^{\mathrm{asym}}(\omega=E_{1})$
at the CdGM-state energy shown in panel (f) are radially symmetric local maxima at all
vortex positions with an oscillating behavior in between vortices.

For an experimental realization, the question about the maximally feasible distance between the tip  ``T'' and contact ``C'' is highly relevant. 
At this point, the local nature of the intrinsic broadening $\Gamma_0$ neglected beyond Eq.~\eqref{eq:rho} will come into play. It violates the assumption of a perfectly coherent subgap state and we expect it to add to the lead-induced broadening $\Gamma$ in the denominator of the non-local conductance $g_{CT}$ of Eq.~\eqref{eq:g_CT} and cause a damping of the non-local conductance peaks. A detailed numerical modeling of the associated crossover to purely local conductance in Eq.~\eqref{eq:rho} would require the addition of a spatially distributed self-energy term in the simulation which is beyond the scope of this paper (and the state-of-the-art). However, we anticipate that the modification of the $g_{CT}$ signal should not compromise the peak-height ratios and the assessment of $q/n$ until the effective broadening reaches the scale $E_0$, compare to the discussion of temperature effects in the leads at the end of Sec.~\ref{sec:Non-local-transport}.

%Another obvious upper limit not built into in our modeling is the decoherence length, caused by electron-electron interaction, phonons, \emph{etc.}. 

Another aspect is possible (single-particle) Anderson localization \citep{EversMirlin:review} in the band of MVM states, which would limit the tip-contact separation to the localization length. However, the two-dimensional Majorana-only model (symmetry class D) is known to feature weak-antilocalization and thus hosts both a localized Anderson insulating phase and a delocalized ``thermal'' metal phase, with a phase diagram that is largely unknown. What has been studied is the transition from a regular triangular Majorana lattice with uniform $\pi/2$ flux through each triangular plaquette (a topological band insulator) to the thermal metal phase which occurs when a randomly chosen minority of $\sim 15$\% of hopping terms have their signs flipped.\citep{Laumann2012} Since the signs of the mutual MVM hopping terms are known to oscillate\citep{Cheng2010} with $k_F^{-1}$ which is on the order of the spread in the intervortex distances, we believe that the realistic systems are well in the thermal metal phase. However, more detailed studies, preferably performed in a Majorana-only effective model\citep{Pathak2021} are desirable.

\section{Conclusion\label{sec:Conclusion}}

We proposed to apply a non-local quantum transport measurement
to identify the presence (or absence) of hybridized Majorana zero
modes in the vortex cores in Fu-Kane materials, like the surface of iron-based superconductors.
In contrast to recent applications of this method to one-dimensional
``Majorana''-wires,\citep{Danon2020,Menard2020} the spatial resolution
inherent in the putative two-dimensional Majorana platforms allows
to extract tell-tale spatial signatures of MVM- or CdGM-states from
the symmetry properties of the peaks in the non-local conductance
trace, see Eq.~\eqref{eq:q/n}. We first treated the case of a vortex
pair analytically and confirmed our findings using transport simulations
based on a lattice model. Finally, we showed that the proposed signatures
persist in the experimentally relevant case of a distorted lattice
model. 

We emphasize that the presented features in the ratio $q/n$ are generic. The only requirement is a sufficiently large hybridization $E_0$, a value that oscillates with separation $R$ in the two-vortex case. If $E_0$ approaches zero as a matter of fine-tuning, the non-local conductance peaks move towards zero-bias and will not be observable such that the experimental protocol cannot be implemented for that particular state. 
This suggests that there is no danger in a false-positive identification of MVM. Further evidence for this also comes from the vortex-lattice case in Fig.~\ref{fig:vortexLattice_transport}(c), where, despite the random (and certainly not fine-tuned) placement of more than ten vortices, the features in question remain clearly distinct. For other perturbations beyond our model (e.g.~disorder potentials), the stability of the proposed signatures remains to be explored.

We expect our results to be relevant for all existing platforms of candidate Fu-Kane materials showing signatures of putative MVMs, see Sec.~\ref{sec:Introduction}. Moreover, our proposal should be applicable to recently suggested alternative realizations of MVMs, like giant topological vortices trapped
in an ordinary superconductors with a dislocation line\citep{Rex2021}.
For future work, it would be interesting to extend our non-local transport
proposal to spin-polarized or superconducting leads\citep{Ruby2015} or to consider the case of a non-negligible intrinsic level broadening.\cite{Liu2017a}

\begin{acknowledgments}
We acknowledge useful discussions with Karsten Flensberg. Computations
were performed at the Lawrencium cluster at Lawrence Berkeley National
Lab. BS, A-PL, MB, RGM, and JEM acknowledge support by the U.S. Department of Energy (DOE), Office of Science, National Quantum Information Science Research Centers., the Quantum Science Center (QSC), a National Quantum Information
Science Research Center of the U.S. Department of Energy (DOE). BS acknowledges financial support by the German National Academy of Sciences Leopoldina through Grant Numbers LPDS 2018-12 and LPDR
2021-01.  MG acknowledges support by the European Research Council (ERC) under the European Union\textquoteright s Horizon 2020 research
and innovation program under grant agreement No.\textasciitilde 856526, and from the Deutsche Forschungsgemeinschaft (DFG) project grant 277101999 within the CRC network TR 183 (subproject C01), and from the Danish National Research Foundation, the Danish Council for Independent Research | Natural Sciences.

\end{acknowledgments}

\onecolumngrid
\appendix

\section{\label{app:Scattering-matrix}Spinful scattering matrix for two normal
leads coupled to a subgap state at energy $E_{0}$}

We start from the general expression of the scattering matrix\cite{Aleiner2002}
\begin{equation}
S(\omega)=1-\frac{2i}{t}H_{LS}\frac{1}{\omega-H_{S}+\frac{i}{t}H_{SL}H_{LS}}H_{SL}\label{eq:S_MahauxWeid}
\end{equation}
which assumes normal (non-superconducting) leads with hopping $t$.
It can be derived from the Fisher-Lee relation which is more complicated
due to one additional matrix inversion. Here, $H_{S}$ is the Hamiltonian
of scattering region to which the leads are coupled with $H_{LS}=H_{SL}^{\dagger}$. 

We now focus on a superconducting system in BdG-formulation and limit
ourselves to a single particle-hole symmetric pair of eigenstates,
$H_{S}\Phi=E_{0}\Phi$ and $H_{S}(P\Phi)=-E_{0}(P\Phi)$. We insert
into Eq.~(\ref{eq:S_MahauxWeid}) and find
\begin{align}
S(\omega) & =1-\frac{2i}{t}W^{\dagger}\frac{1}{\left(\begin{array}{cc}
\omega-E_{0} & 0\\
0 & \omega+E_{0}
\end{array}\right)+\frac{i}{t}WW^{\dagger}}W,\label{eq:S_MahauxWeid_SC}\\
W & \equiv\left(\begin{array}{c}
\Phi^{\dagger}\\
\left(P\Phi\right)^{\dagger}
\end{array}\right)H_{SL}.
\end{align}
We further assume a set of leads such that $H_{SL}$ is diagonal in
the lead index $\alpha$. For lead $\alpha$, we have in the BdG-formulation
$H_{SL}^{\alpha}=\gamma_{\alpha}\tau_{z}$ with $\gamma_{\alpha}\in\mathbb{R}$
a spin-independent hopping. 

For the spinless case, we can chose $\Phi=(u^{\star},v)^{\mathrm{T}}$
and with $\mathcal{P}=\tau_{x}\mathcal{K}$ we find
\begin{equation}
W_{\alpha}=\gamma_{\alpha}\left(\begin{array}{cc}
u_{\alpha} & -v_{\alpha}^{\star}\\
v_{\alpha} & -u_{\alpha}^{\star}
\end{array}\right)
\end{equation}
where $u_{\alpha}=u(\mathbf{r}_{\alpha})$ is the electron part of
the BdG-wavefunction at the position of lead $\alpha$ and similar
for the hole-part $v_{\alpha}$. For the spinless case and in the
presence of two leads $\alpha=\{L,R\}$, Ref.~\cite{Danon2020} derived
an explicit expression for the scattering matrix and conductances.
\begin{table}
\noindent \begin{centering}
\begin{tabular}{|c|c|c|c|}
\hline 
$|u_{\alpha}|^{2}\equiv\sum_{\sigma}|u_{\alpha,\sigma}|^{2}$ & $n_{\alpha}\equiv|u_{\alpha}|^{2}+|v_{\alpha}|^{2}$ & $\Gamma_{\alpha}\equiv\frac{\gamma_{\alpha}^{2}}{t}n_{\alpha}$ & $\Gamma\equiv\Gamma_{L}+\Gamma_{R}$\tabularnewline
\hline 
$|v_{\alpha}|^{2}\equiv\sum_{\sigma}|v_{\alpha,\sigma}|^{2}$ & $q_{\alpha}\equiv|u_{\alpha}|^{2}-|v_{\alpha}|^{2}$ & $\xi_{\alpha}\equiv\frac{\gamma_{\alpha}^{2}}{t}q_{\alpha}$ & $\xi^{2}\equiv\xi_{L}^{2}+\xi_{R}^{2}$\tabularnewline
\hline 
$\left[uv\right]_{\alpha}\equiv\sum_{\sigma}u_{\sigma\alpha}v_{\sigma\alpha}$ & $\Xi_{\alpha}\equiv\gamma_{\alpha}^{4}\left|\left[uv\right]_{\alpha}\right|^{2}$ & $4|u_{\alpha}|^{2}|v_{\alpha}|^{2}=n_{\alpha}^{2}-q_{\alpha}^{2}$ & $\xi_{LR}^{2}=\frac{1}{t^{2}}\gamma_{L}^{2}\gamma_{R}^{2}\left(\left[uv\right]_{R}\right)^{\star}\left[uv\right]_{L}$\tabularnewline
\hline 
$a\equiv2\gamma_{L}^{2}\left[uv\right]_{L}+2\gamma_{R}^{2}\left[uv\right]_{R}$ & $\left|a\right|^{2}=8\mathrm{Re}\xi_{LR}^{2}+4\Xi_{L}+4\Xi_{R}$ & $b_{\pm}\equiv\omega\pm E_{0}+i\Gamma$ & $c\equiv\omega^{2}-\Gamma^{2}-E_{0}^{2}+|a|^{2}$\tabularnewline
\hline 
\end{tabular}
\par\end{centering}
\caption{\label{tab:sum}Summary of abbreviations used in the analytical calculation in App.~\ref{app:Scattering-matrix}}
\end{table}

We now generalize the calculation for the spinful case where $\Phi=(u_{\uparrow},u_{\downarrow},v_{\downarrow},-v_{\uparrow})$
and $\mathcal{P}=\sigma_{y}\tau_{y}\mathcal{K}$. In this case, we
obtain
\begin{equation}
W_{\alpha}=\gamma_{\alpha}\left(\begin{array}{cccc}
u_{\uparrow,\alpha}^{\star} & u_{\downarrow,\alpha}^{\star} & -v_{\downarrow,\alpha}^{\star} & v_{\uparrow,\alpha}^{\star}\\
v_{\uparrow,\alpha} & v_{\downarrow,\alpha} & -u_{\downarrow,\alpha} & u_{\uparrow,\alpha}
\end{array}\right).
\end{equation}
We set $t\equiv1$ in the following and use the definitions and relations
in Table \ref{tab:sum} some of which already appeared in the main
text. We find

\begin{equation}
iWW^{\dagger}=i\left(\begin{array}{cc}
\Gamma & a^{\star}\\
a & \Gamma
\end{array}\right)
\end{equation}
and insert this in Eq.~(\ref{eq:S_MahauxWeid_SC}) where $\alpha,\beta=\{L,R\}$.
\begin{eqnarray}
S_{\alpha\beta}(\omega) & = & \delta_{\alpha\beta}-\frac{2i\gamma_{\alpha}\gamma_{\beta}}{c+2i\Gamma\omega}\left(\begin{array}{cc}
u_{\uparrow,\alpha} & v_{\uparrow,\alpha}^{\star}\\
u_{\downarrow,\alpha} & v_{\downarrow,\alpha}^{\star}\\
-v_{\downarrow,\alpha} & -u_{\downarrow,\alpha}^{\star}\\
v_{\uparrow,\alpha} & u_{\uparrow,\alpha}^{\star}
\end{array}\right)\left(\begin{array}{cc}
b_{+} & -ia^{\star}\\
-ia & b_{-}
\end{array}\right)\left(\begin{array}{cccc}
u_{\uparrow,\beta}^{\star} & u_{\downarrow,\beta}^{\star} & -v_{\downarrow,\beta}^{\star} & v_{\uparrow,\beta}^{\star}\\
v_{\uparrow,\beta} & v_{\downarrow,\beta} & -u_{\downarrow,\beta} & u_{\uparrow,\beta}
\end{array}\right).
\end{eqnarray}
We now extract the sub-matrices required for computing the local-
and non-local conductance, $g_{LL}$ and $g_{LR}$.
\begin{align}
g_{LL} & =N_{L}-\mathrm{tr}\left[s_{ee,LL}^{\dagger}s_{ee,LL}\right]+\mathrm{tr}\left[s_{he,LL}^{\dagger}s_{he,LL}\right]\\
g_{LR} & =-\mathrm{tr}\left[s_{ee,LR}^{\dagger}s_{ee,LR}\right]+\mathrm{tr}\left[s_{he,LR}^{\dagger}s_{he,LR}\right]
\end{align}
After straightforward but lengthy algebra, we obtain 
\begin{eqnarray}
g_{LL}(\omega) & = & \frac{4}{\left|c+2i\Gamma\omega\right|^{2}}\left[4c\left(\Xi_{L}+\mathrm{Re}\left[\xi_{LR}^{2}\right]\right)+\left(\Gamma\Gamma_{L}-\xi_{L}^{2}\right)\left(2\omega^{2}-c\right)+\omega\xi_{L}\left\{ 2\Gamma_{R}E_{0}-8\mathrm{Im}\left[\xi_{LR}^{2}\right]\right\} \right],\label{eq:g_LL_full}\\
g_{LR}(\omega) & = & \frac{4\xi_{L}}{\left|c-2i\Gamma\omega\right|^{2}}\left\{ \xi_{R}\left(c-2\omega^{2}\right)-\omega\left(2\Gamma_{R}E_{0}-8\mathrm{Im}\xi_{LR}^{2}\right)\right\} .\label{eq:g_LR_full}
\end{eqnarray}
In the main text, we are only interested in $g_{LR}(\omega)$. We
obtain Eq.~\eqref{eq:g_CT} for $|\omega|\simeq E_{0}$ assuming that $E_{0}$
is much larger than all other scales appearing in Eq.~(\ref{eq:g_LR_full}). 

\twocolumngrid
\bibliography{library}

\end{document}